\documentclass[12pt]{iopart}

\pdfoutput=1
 \usepackage[square,numbers,sort&compress]{natbib}
\usepackage[breaklinks=true,colorlinks=true,linkcolor=blue,urlcolor=blue,citecolor=blue]{hyperref}
\usepackage{latexsym}
\usepackage{graphics}
\usepackage{graphicx}
\usepackage{epsfig}
\usepackage{mathcomp}
\usepackage{amsopn}
\usepackage{amsfonts}
\usepackage{amstext}
\usepackage{amssymb}
\usepackage{amstext}
\usepackage{subfigure}
\usepackage{braket}
\usepackage{iopams}  
\usepackage{wasysym}
\usepackage{iopams}

\usepackage[makeroom]{cancel}
\usepackage{epstopdf}
\usepackage{braket}
\newcommand{\levicivita}{}% initialize
\def\levicivita#1#{\tensor#1{\epsilon}}
\begin{document}
\title []{Benchmarking discrete truncated Wigner approximation and neural network quantum states with the exact dynamics in a Rydberg atomic chain}
%\title{Analayzing excitation and R\'enyi entropy dynamics in a Rydberg atomic chain via discrete truncated Wigner approximation and artificial neural networks}
\author{Vighnesh Naik$^{1}$\footnote{\label{n1}These authors contributed equally to this work.}, Varna Shenoy K$^{1}$\footref{n1}, Weibin Li$^2$, and Rejish Nath$^1$}
\address{$^1$Department of Physics, Indian Institute of Science Education and Research Pune, Dr. Homi Bhabha Road, Pune- 411008, Maharashtra, India. \\
$^2$ School of Physics and Astronomy, and Centre for the Mathematics and Theoretical Physics of Quantum Non-Equilibrium Systems, University of Nottingham, NG7 2RD, United Kingdom.}
\ead{rejish@iiserpune.ac.in}
\vspace{10pt}
%\begin{indented}
%\item[]February 2014
%\end{indented}

\begin{abstract}
We benchmark the discrete truncated Wigner approximation (DTWA) and Neural quantum states (NQS) based on restricted Boltzmann-like machines with the exact excitation and correlation dynamics in a chain of ten Rydberg atoms. The initial state is where all atoms are in their electronic ground state. We characterize the excitation dynamics using the maximum and time-averaged number of Rydberg excitations. DTWA results are different from the exact dynamics for large Rydberg-Rydberg interactions. In contrast, by increasing the number of hidden spins, the NQS can be improved but still limited to short-time dynamics. Interestingly, irrespective of interaction strengths, the time-averaged number of excitations obtained using NQS is in excellent agreement with the exact results. Concerning the calculation of quantum correlations, for instance, second-order bipartite and average two-site R\'enyi entropies, NQS looks more promising. Finally, we discuss the existence of a power law scaling for the initial growth of average two-site R\'enyi entropy.
\end{abstract}

%\submitto{\JPB}
\maketitle
%\tableofcontents
\section{Introduction}
In general, analyzing the dynamics of a quantum many-body system is a formidable task because of the exponentially large Hilbert space. Further, strong quantum correlations or entanglement growth makes quantum dynamics more intriguing. Developing numerical approaches accurately capturing the quantum correlations, including their dynamical growth, is an ongoing activity \cite{czi20}. It led to the development of various numerical techniques based on tensor network states and phase-space methods. The former includes, for instance, density-matrix renormalization group (DMRG) (for both ground states and dynamics) \cite{oru14, whi92, vid04, sch11, bri17,whi04} in which the Hilbert space is truncated to states with significant probabilities, leading to a polynomial growth in Hilbert space with system size. DMRG is an excellent tool for studying one-dimensional systems with weak entanglement. The phase-space methods are semi-classical and involve solving classical equations of motion for the phase-space variables. Examples are the truncated Wigner approximation (TWA) \cite{pol10, bla08} and its discrete version (DTWA) \cite{sch15, scha15,puc16}. DTWA was initially developed for spin-1/2 systems and later generalized to higher spin \cite{zhu19}, higher dimensional phase space \cite{wur18}, and dissipative systems \cite{hub21}. DTWA is also employed to study dynamical phase transitions in large spin systems \cite{kha20}. The results from DTWA have shown good agreement with experiments of Rydberg \cite{ori18, sig21, gei21, hao21} and dipolar atoms \cite{lep21, fer19, pat20} where an XY spin-model describes the system. Generally, such a good agreement is only sometimes guaranteed and is limited to a few initial cycles, depending on the strength and range of inter-particle interactions \cite{kun21}.

At the same time, there is a growing interest in using machine learning methods to study physics problems \cite{meh19, car19}. Artificial neural networks are proposed to explore quantum many-body systems in which the quantum states are represented by restricted-Boltzmann-machine (RBM)-like network architecture but with complex weights, and are called neural network quantum states \cite{car17,fab19, czi18, wuy20,den17a,den17b, ann22}. Due to its design, neural network quantum states (NQS) are naturally suited for spin-1/2 systems, which have shown good agreement in predicting ground states. Whereas for dynamics, it is limited for short periods \cite{fab19, czi18, wuy20}. 

On the other side, quantum simulators based on ultra-cold Rydberg atoms have shown tremendous progress, extending to large system sizes beyond a regime where classical computers can tackle \cite{bro20, ber17, kim18, gua18,lab16, gra19, scho20,schl20, eba20, blu21}. Hence, it becomes necessary to test advanced numerical approaches against the exact dynamics of atomic lattices with Rydberg excitations. Rydberg atoms are known to exhibit prodigious long-range interactions \cite{saf10, beg13} leading to the phenomenon of the Rydberg, or dipole blockade \cite{luk01,gae09, urb09}. Blockade and anti-blockade can lead to strong correlation effects, which can simulate non-trivial phases in condensed-matter \cite{ber17, wei10, ric11,sch12,bar15,zei16,zei17,mar17,gro17} and find applications in quantum information protocols \cite{saf10,jak00,wil10,ise10,saf16, sls20, xia21}. Typically, the Rydberg atomic setup has been modeled as a gas of interacting two-level atoms, with either dipolar or van der Waals-type interactions \cite{gla12}. Further, making the atom-light couplings time-dependent provides fine-tuning in quantum state preparation and expands the territory of problems that can be addressed \cite{blu21, bas18, nir20, mal21, dhi23}.

In this work, we numerically analyze the excitation and correlation dynamics in a small chain of ten atoms where each atom is initially in its ground state, coupled to a Rydberg state by a light field. In particular, we explore the oscillatory behavior of the number of excitations, the dependence of the maximum and the average number of excitations on the interaction strengths, and finally, the dynamics of bipartite and mean two-site second-order R\'enyi entropy. The maximum number of excitations is always seen during the first oscillation cycle. The Rydberg blockade suppresses the excitation fraction for significant interaction strengths, leading to a saturation in the average and maximum number of excitations. For sufficiently small interaction strengths, collapse and revival exist in the dynamics of the total number of excitations at longer times. Interestingly, a power law scaling for the initial growth of the mean two-site second-order R\'enyi entropy exists, with the scaling exponent depending enormously on the interaction strengths.

Our work compares the exact results with those obtained via DTWA and neural network quantum states. Since the number of Rydberg qubits in experiments has increased tremendously over the last couple of years,  testing various numerical methods against the exact results of small systems is necessary. The latter may help identify ways to improve the existing techniques or even lead to the development of entirely new approaches, eventually for studying strongly correlated systems.

In DTWA, we show results with and without incorporating the second-order correlations from the Bogoliubov–Born–Green–Kirkwood–Yvon (BBGKY) hierarchy,  termed first and second-order DTWA, respectively. DTWA and network quantum states accurately capture the excitation dynamics over significant periods for small interaction strengths. However, the revival of the excitation population is not seen in both methods. As interaction strengths increase, the agreement with the exact results from both methods starts to deviate at longer times. However, the network quantum state is more reliable, especially when computing the average and maximum number of excitations. The initial power-law growth in the mean two-site second-order R\'enyi entropy is accurately predicted by NQS, even for larger interactions. The second-order DTWA exhibits numerical instabilities for considerable interaction strengths, also reported in the study of quantum spin models \cite{kun21}.  
%bipartite and mean two-site second-order R\'enyi entropy

The paper is structured as follows. In section \ref{sup}, we introduce the physical setup and the governing Hamiltonian. In sections~\ref{DTWA} and \ref{ann}, the details of the DTWA and NQS techniques are provided. The excitation and correlation dynamics are shown in sections~\ref{ssd} and \ref{re}, respectively. Finally, we provide conclusions and outlook in section~\ref{con}.

%%%%%%%%%

\section{Setup and Model}
\label{sup}
%%%%%

%%%%%%%%%%%%%%%%%%%%%%%%
%%Figure 0
%%%%%%%%%%%%%%%%%%%%%%%%
\begin{figure}
\centering
\includegraphics[width=0.7 \columnwidth]{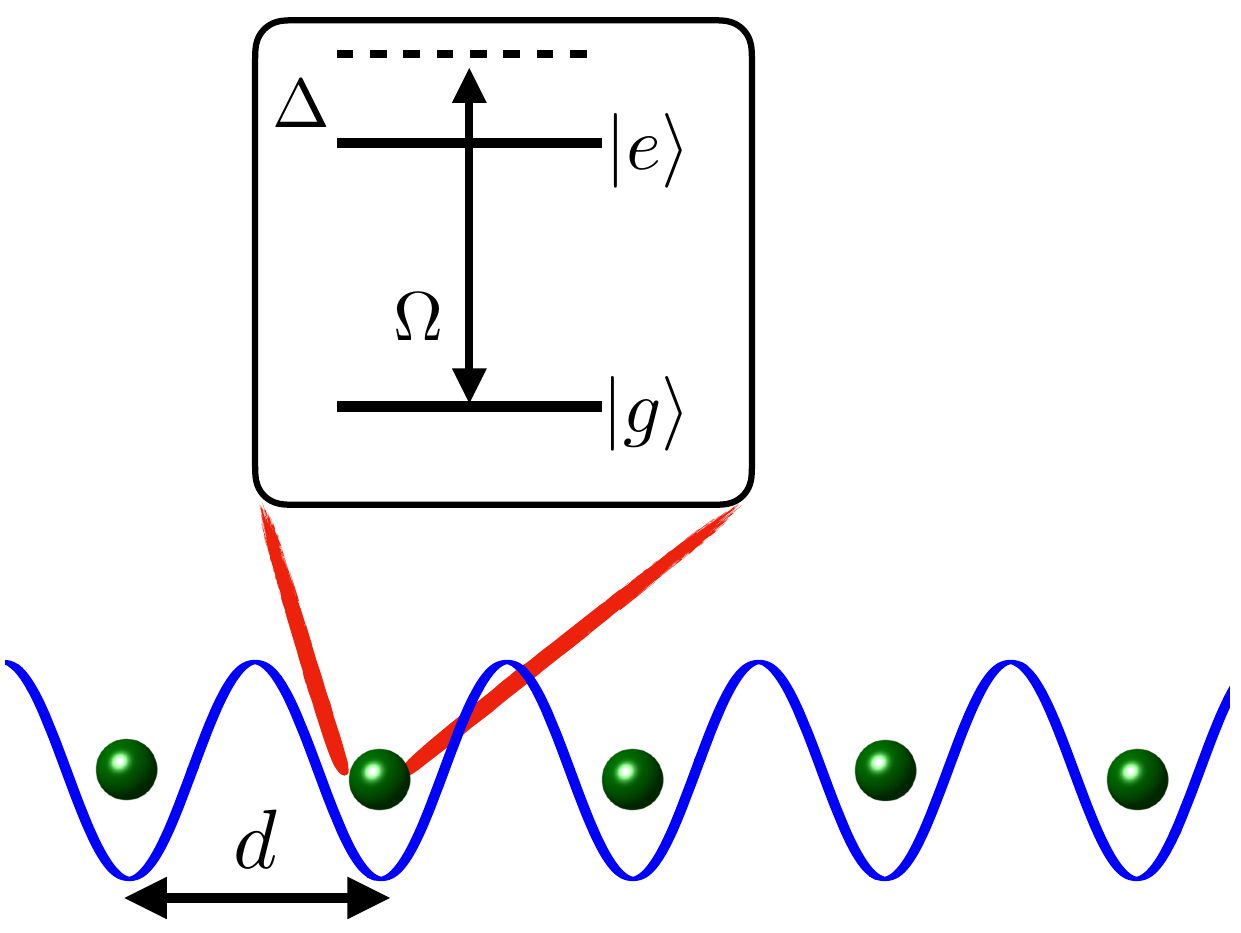}
\caption{The schematic diagram of the setup consists of an array of atoms with a spacing of $d$. For each atom, the ground state $|g\rangle$ is coupled to a Rydberg state $|e\rangle$ with a light field with a detuning $\Delta$ and Rabi frequency $\Omega$.}
\label{fig:0}
\end{figure}
%%%%%%%
We consider a chain of atoms with lattice spacing $d$ as shown in Figure~\ref{fig:0}. The electronic ground state $|g\rangle$ of each atom is coupled to a Rydberg state $|e\rangle$ with a detuning $\Delta$ and a Rabi frequency $\Omega$. In the frozen gas limit, the system is described by the Hamiltonian ($\hbar=1$):
\begin{equation}
\hat H(t)=-\Delta\sum_{i=1}^N\hat\sigma_{ee}^{i}+\frac{\Omega}{2}\sum_{i=1}^N\hat\sigma_x^{i}+\sum_{i<j}V_{ij}\hat\sigma_{ee}^{i}\hat\sigma_{ee}^{j},
\label{ham}
\end{equation}
where $\hat\sigma_{ab}=|a\rangle\langle b|$ with $a, b\in \{e, g\}$ includes both transition and projection operators, and $\hat\sigma_x=\hat\sigma_{eg}+\hat\sigma_{ge}$. We assume the Rydberg excited atoms interact via van der Waals potential $V_{ij}=C_6/r_{ij}^6$ \cite{beg13}, where $C_6$ is the interaction coefficient, and $r_{ij}$ is the separation between $i$th and $j$th Rydberg excitations. The exact dynamics of the system is analyzed by solving the Schr\"odinger equation: $i\partial \psi(t)/\partial t=\hat H(t)\psi(t)$. 

Introducing $\hat\sigma_{ee}^i=(\mathcal I+\hat\sigma_{z}^i)/2$ where $\mathcal I$ is the identity matrix, we rewrite the Hamiltonian as, $\hat H=\sum_i \hat H_i+\sum_{i<j}\hat H_{ij}$, where $\hat H_i=-\Delta^{{\rm eff}}_i\hat\sigma_{z}^{i}+\Omega\hat\sigma_x^{i}/2$ consists of single particle terms with an effective detuning $\Delta^{{\rm eff}}_i=\Delta-\sum_{j\neq i}V_{ij}/2$, and $\hat H_{ij}=V_{ij}\hat\sigma_{z}^{i}\hat\sigma_{z}^{j}/4$ contains the two-body interactions. Henceforth, we take $\Delta=0$. The model is then a quantum Ising model with an additional longitudinal field and long-range Ising interactions. The latter makes this model different from previous studies on quantum Ising model using DTWA and neural quantum state. We characterize the dynamics via experimentally relevant quantities such as the instantaneous number of excitations $N_e(t)=\sum_i\langle\psi(t)|\hat\sigma_{ee}^i|\psi(t)\rangle$, the maximum $N_e^{max}$ and the average number of excitations $N_e^{avg}=(1/T)\int_0^TN_e(t)$ over a period of time $T$. Later in section~\ref{re}, we analyze bipartite and mean two-site second-order R\'enyi entropy of the corresponding dynamics.

%%%%%%%%%%%

\section{Discrete truncated Wigner approximation: Equations of motion}
\label{DTWA}
Here, we briefly outline the method of DTWA. Each atom forms a two-level system (equivalently a qubit or a spin-1/2 particle) composed of the ground state $|g\rangle$ and Rydberg state $|e\rangle$. The finite and discrete Hilbert space is mapped onto a discrete quantum phase-space via the Wigner-Weyl transform. A single qubit discrete phase-space is defined as a real-valued finite field spanned by four-phase points, denoted by $\alpha\equiv (m, n)\in\{(0, 0), (0, 1), (1, 0), (1, 1)\}$ \cite{woo87,woo03, sch15}. For each $\alpha$, one can associate a phase-point operator $\hat A_{\alpha}=(\mathcal I+{\bf r}_\alpha\cdot\hat{\bf \sigma})/2$, where $\mathcal I$ is the identity operator and  $\hat{\bf \sigma}=(\hat\sigma_x, \hat\sigma_y, \hat\sigma_z)$ are the Pauli matrices. The three-vectors $\{{\bf r}_\alpha\}$ do not have a unique choice, and the best suitable sampling scheme for a given Hamiltonian can be appropriately selected \cite{puc16}. For the initial state we take, the best results are obtained using ${\bf r}_{(1,0)}=(1, 0, -1)$, ${\bf r}_{(1,1)}=(-1,0, -1)$, ${\bf  r'}_{(1,0)}=(0, 1, -1)$ and ${\bf  r'}_{(1,1)}=(0,-1, -1)$, which is kept throughout. 

Any observable $\hat \mathcal O$ from the Hilbert space is mapped into a Weyl symbol $\mathcal O^W_{\alpha}=\Tr(\hat\mathcal O\hat A_{\alpha})/2$ in the discrete phase-space. The density operator $\hat\rho$ is written as $\hat\rho=\sum_\alpha w_\alpha\hat A_{\alpha}$, where the weights $w_\alpha=\Tr(\hat\rho\hat A_{\alpha})/2$ form a quasi-probability distribution similar to the original Wigner function and is the Weyl symbol of the density matrix. $w_\alpha$ can also take negative values. For a system of $N$ two-level atoms, we have a discrete phase-space of $4^N$ points denoted simply by $\vec \alpha=\{\alpha_1, \alpha_2, ...,\alpha_N\}$. Let the initial density matrix be, 
\begin{equation}
\hat{\rho}_0 = \sum_{\vec \alpha} W_{\vec \alpha} \  \hat{A}_{\alpha_1}\otimes\hat{A}_{\alpha_2}\otimes...\otimes\hat{A}_{\alpha_N},
\label{idm}
\end{equation}
which corresponds to a product state with $W_{\vec \alpha}=w_{\alpha_1}w_{\alpha_2}...w_{\alpha_N}$ and $\sum_{\vec \alpha}=\sum_{\alpha_1, \alpha_2, ...}$. The density matrix at time $t$ is defined as
	\begin{equation}
	\hat{\rho}(t) = \sum_{\vec \alpha} W_{\vec \alpha} \  \hat{\mathcal A}^{\vec \alpha}_{1...N}(t),
	\end{equation}
	with 
	\begin{equation}
		\hat{\mathcal A}^{\vec \alpha}_{1...N}(t) = \hat U(t)\hat{A}_{\alpha_1}\otimes\hat{A}_{\alpha_2}\otimes.....\otimes\hat{A}_{\alpha_N}  \hat U^\dagger(t)
	\end{equation}
	where $ \hat U(t)=\exp(-i\hat Ht)$ is the unitary time evolution operator. Here, we do not consider all the trajectories but do a monte carlo sampling of $N_s$ number of trajectories, and hence the density matrix is defined as,
\begin{equation}
    \hat{\rho}(t) = \sum_{\alpha} \frac{1}{N_s}\hat{\mathcal{A}}^{\vec \alpha}_{1...N}(t).
    \label{dm2}
\end{equation} 
The operator $\hat{\mathcal A}^{\vec \alpha}_{1...N}(t)$ satisfies the Liouville-von Neumann equation \cite{puc16}:
	\begin{equation}
	    i \frac{\partial}{\partial t}\hat{\mathcal A}^{\vec \alpha}_{1...N} = \left[\hat H,\hat{\mathcal A}^{\vec \alpha}_{1...N}\right]
	\end{equation}
	We can treat the operator $\hat{\mathcal A}^{\vec \alpha}_{1...N}$ as a quasi-density-matrix since its trace is equal to one, but need not be a positive definite. The reduced $\hat {\mathcal A}$ operators are obtained by tracing out remaining parts of the system as,
	\begin{equation}
		\hat{\mathcal A}^{\vec \alpha}_i = \mathrm{Tr}_{\cancel i}\hat{\mathcal A}^{\vec \alpha}_{1...N}, \ \ \ \ \ \ \ \ \ \ \ \ \ \ \hat{\mathcal A}^{\vec \alpha}_{ij} = \mathrm{Tr}_{\cancel i\cancel j}\hat{\mathcal A}^{\vec \alpha}_{1...N} \ 
	\end{equation}
	where $\mathrm{Tr}_{\cancel i}$ denotes a partical trace over all the indices except $i$ and similarly for $\mathrm{Tr}_{\cancel i\cancel j}$ with $i\neq j$. For reduced density operators, using the Liouville-von Neumann equation, one can write down the hierarchy of equations of motions, so-called the BBGKY hierarchy. Following the same prescription, a similar hierarchy of equations of motions for the reduced $\hat{\mathcal A}$ operators is obtained, by introducing the cluster expansion \cite{puc16},
	\begin{eqnarray}
	\hat{\mathcal A}_{ij} & = \hat{\mathcal A}_i\hat{\mathcal A}_j + \hat{\mathcal C}_{ij}\\
	\hat{\mathcal A}_{ijk} & = \hat{\mathcal A}_i\hat{\mathcal A}_j\hat{\mathcal A}_k + \hat{\mathcal A}_i \hat{\mathcal C}_{jk}+\hat{\mathcal A}_j  \hat{\mathcal C}_{ik}+ .... + \hat{\mathcal C}_{ijk}\\
	\hat{\mathcal A}_{ijkl} & = \hat{\mathcal A}_i\hat{\mathcal A}_j\hat{\mathcal A}_k\hat{\mathcal A}_l + \hat{\mathcal A}_i\hat{\mathcal A}_j \hat{\mathcal C}_{kl}+...+\hat{\mathcal{C}}_{ij} \hat{\mathcal{C}}_{kl} +... +\hat{\mathcal C}_{ijk}\hat{\mathcal A}_l + .... + \hat{\mathcal C}_{ijkl},
	\end{eqnarray}
	where $\hat{\mathcal A}_1,\hat{\mathcal A}_2,...,\hat{\mathcal A}_N$ are the uncorrelated parts of $\hat{\mathcal A}_{1...N}$, and $\hat{\mathcal C}_{ij},\hat{\mathcal C}_{ijk}...$ operators incorporate the correlations between the particles arising from the inter-particle interactions.  Note that, we have removed the superscript $\vec \alpha$ in $\hat{\mathcal A}_j$ for simplicity. Truncating beyond two-particle correlations, one obtains the first two equations of the BBGKY hierarchy as:
    \begin{eqnarray}\label{BBGKY}
    i \frac{\partial}{\partial t} \hat{\mathcal A}_i& = & [\hat H_i, \hat{\mathcal A}_i] + \sum_{k\neq i}\mathrm{Tr}[\hat H_{ik}, \hat{\mathcal C}_{ik} + \hat{\mathcal A}_i \hat{\mathcal A}_k] \\
    i \frac{\partial}{\partial t} \hat{\mathcal C}_{ij}& = & [\hat H_i + \hat H_j + \hat H^{\mathrm{H}}_{i\cancel j} + \hat H^{\mathrm{H}}_{j\cancel i}, \hat{\mathcal C}_{ij}] + [\hat H_{ij}, \hat{\mathcal C}_{ij} + \hat{\mathcal A}_i\hat{\mathcal A}_j] +\nonumber\\
    && \sum_{k\neq i,j}\left(\mathrm{Tr}_k[\hat H_{ik},\hat{A}_i\hat{\mathcal C}_{jk}] + \mathrm{Tr}_k[\hat H_{jk},\hat{\mathcal A}_j\hat{\mathcal C}_{jk}]\right)-\nonumber \\
    && \hat{\mathcal A}_i\mathrm{Tr}_i[\hat H_{ij},\hat{\mathcal C}_{ij} + \hat{\mathcal A}_i\hat{\mathcal A}_j] - \hat{\mathcal A}_j\mathrm{Tr}_j[\hat H_{ij},\hat{\mathcal C}_{ij} + \hat{\mathcal A}_i\hat{\mathcal A}_j]     \end{eqnarray}
    where $\hat H^\mathrm{H}_{i \cancel j}$ is a Hartree operator or Mean-field operator given by:
    \begin{equation}
    	\hat H^\mathrm{H}_{i \cancel j} = \sum_{k\neq i,j} \mathrm{Tr}_k(\hat H_{ik}\hat{\mathcal A}_k).
    \end{equation}
At this point, one expands $\mathcal A$ and $\mathcal C$ operators in the basis of Pauli spin matrices,
 \begin{eqnarray}
\hat{\mathcal A}_i & = \frac{1}{2}(\mathcal I + \hat a_i\cdot\hat\sigma)\\
\hat{\mathcal C}_{ij} & = \frac{1}{4}\sum_{\mu,\nu \in \{x,y,z\}} c^{\mu\nu}_{ij}\hat\sigma_\mu^i \hat\sigma_\nu^j,
\end{eqnarray}
for $i\neq j$. Finally, one obtains the equation of motion for $a_i^\mu$ as, 
\begin{eqnarray}
\label{dte1}
	 	\frac{1}{2}\dot{a}^{\mu}_i& = & \sum_{\gamma} \left[\frac{\Omega}{2} a^\gamma_i \varepsilon^{\mu x \gamma}-\frac{ \Delta^{\mathrm{eff}}_i}{2} a^\gamma_i\varepsilon^{\mu z \gamma} + \left(G^z_{i} a^\gamma_i + G^{z\gamma}_{i}\right)\varepsilon^{\mu z\gamma}\right],
	\end{eqnarray}
and that of the two-body correlations are 
\begin{equation}
\frac{1}{2}\dot{c}^{\mu\nu}_{ij}=T_1+T_2+T_3+T_4+T_5
\label{dte2}
\end{equation}
 with
\begin{eqnarray}
	 	T_1 &= & \ \sum_{\beta}\frac{V_{ij}}{4}\left(a^\beta_i  \delta_{\nu z} - a^\beta_j \delta_{\mu z}\right)\varepsilon^{\mu\nu\beta}, \\
		T_2&=& \sum_{\delta} c^{\delta\nu}_{ij}\left[\frac{\Omega}{2} \varepsilon^{x\delta\mu} + \left(G^z_{i\cancel{j}}-\frac{\Delta^{\mathrm{eff}}_i}{2}\right) \varepsilon^{z\delta\mu}\right],\\
	 	T_3&=&\sum_{\gamma} c^{\mu\gamma}_{ij} \left[\frac{\Omega}{2} \varepsilon^{x\gamma\nu} + \left( G^z_{j\cancel{i}}-\frac{\Delta^{\mathrm{eff}}_j}{2} \right) \varepsilon^{z\gamma\nu}\right], \\
		T_4&=& \sum_{\gamma} \left[G^{\nu z}_{ij}a^\gamma_i \varepsilon^{z\gamma\mu}+ G^{\mu z}_{ji} a^\gamma_j \varepsilon^{z\gamma\nu}\right], \\
	 	T_5&=& - \sum_{\gamma} \frac{V_{ij}}{4}\left[a^\mu_i\left(c^{z\gamma}_{ij} + a^z_i a^\gamma_j\right)\varepsilon^{z\gamma\nu} + a^\nu_j\left(c^{\gamma z}_{ij} + a^\gamma_i a^z_j\right)\varepsilon^{z\gamma\mu}\right],
		\label{t5}
\end{eqnarray}
where $\varepsilon$ is the Levi-Civita symbol, $G^z_{i} = \sum_{k\neq i}V_{ik} a^z_k/4$, $G^{z\gamma}_{i} = \sum_{k\neq i}V_{ik}c^{z\gamma}_{ki}/4$, $G^z_{i\cancel{j}} = \sum_{k\neq i,j}V_{ik}a^z_k/4$, and $G^{\nu z}_{ij} = \sum_{k\neq i,j}V_{ik}c^{\nu z}_{jk}/4$. As we found, not all $T_j$ terms may become relevant in the dynamics. Further, the terms proportional to the cube of $a^\gamma_j$ in equation~(\ref{t5}) and $T_4$ lead to numerical instabilities even for small interaction strengths and are omitted.

Equations~(\ref{dte1}) and (\ref{dte2}) describe the dynamics of a Rydberg atomic chain in DTWA incorporating the two-body correlations using the BBGKY heirarchy. Neglecting $c^{\mu\nu}_{ij}$, we get the results at the mean-field level (except the fluctuations in the initial state from the sampling) or so-called the first-order DTWA. In second-order DTWA, the correlations $c^{\mu\nu}_{ij}$ are incorporated in the dynamics. Equations~(\ref{dte1}) and (\ref{dte2}) are solved using Runge-Kutta method. Once $a_i^\mu$ are obtained, the expectation value of any single particle operator can be calculated as,
    	\begin{eqnarray}
    	\langle\hat\sigma_\mu^i\rangle(t) =\mathrm{Tr}\left(\hat\sigma_\mu^i\hat{\rho}(t)\right) = \sum_\alpha W_\alpha \mathrm{Tr}\left(\hat\sigma_\mu^i\hat{\mathcal A}^{\vec\alpha}_{1....N}(t)\right) = \sum_\alpha W_\alpha a^\mu_i(t).
	\label{exn}
    	\end{eqnarray}
The excitation number $N_e(t)$ is obtained by computing $\langle\sigma_z^i\rangle(t)$ or equivalently $a^z_i(t)$. The summation over $\alpha$ in equation~(\ref{exn}) is over all classical trajectories. Since, the number of such trajectories grow exponentially with system size, we use a Monte Carlo sampling with $W_\alpha$ as the probability distribution. For $N=10$, there are $4^{10}$ possible trajectories and we use $N_s=20000$ trajectories for which the dynamics is already converged.

%%%%%%%%%	
%%%%%

%%%%%%%%%%%%%%%%%%%%%%%%
%%Figure 0a
%%%%%%%%%%%%%%%%%%%%%%%%
\begin{figure}
\centering
\includegraphics[width=0.75 \columnwidth]{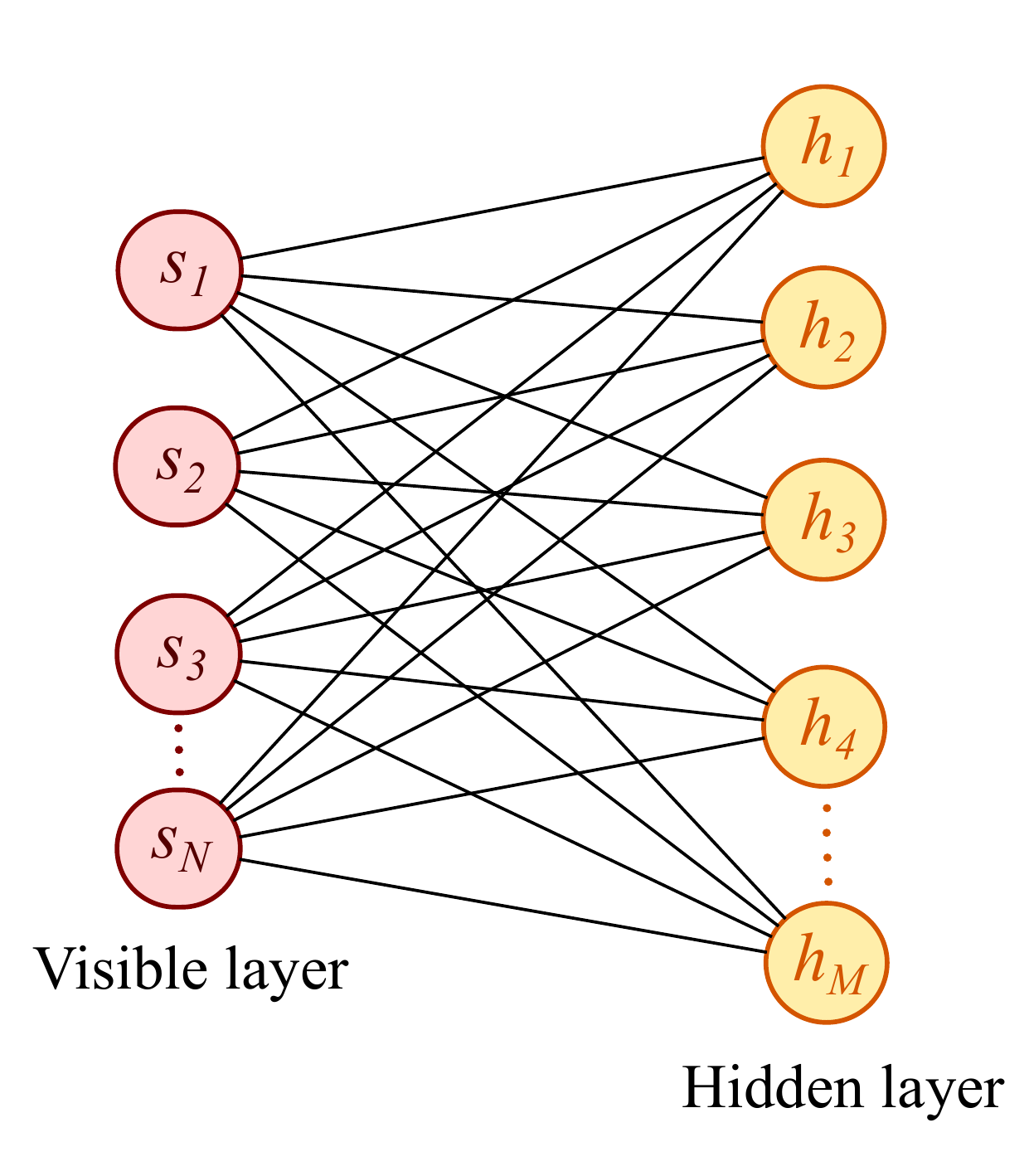}
\caption{The schematic diagram of the neural network quantum states, which is identical to RBM-like architecture. Each physical spin 
$s_i$ in the visible layer is coupled to every spin $h_i$ in the hidden or auxiliary layer.}
\label{fig:0a}
\end{figure}

%%%%%
\section{Neural network quantum states}
\label{ann}
This section briefly describes the NQS based on the RBM-like architecture, schematically shown in figure~\ref{fig:0a} \cite{car17,fab19, czi18, den17a, den17b}.  The physical spins (neurons) $s_i\in \{1, -1\}$ in the primary layer are complemented by an auxiliary layer of Ising spins (neurons) $h_j\in \{1, -1\}$. Each node in the primary layer is connected to every node in the auxiliary layer. There are no connections between nodes within a given layer. The connection between $s_i$ and $h_j$ has a complex weight $W_{ij}$ which is $(ij)^{th}$ element of the weight matrix $\textbf{\emph{W}}$ and $b_j$ are the bias weights for the hidden units with $i=1, ..., N$ and $j=1, ..., M$. We define $M=\gamma N$. Now, the neural many-body quantum state (unnormalized) is defined as $ \ket{\Psi} = \sum_{S} \Psi(S, {\bf x}) \ket{S}$ with variational ansatz
\begin{equation}
        \log(\Psi(S, {\bf x})) = \sum_{j=1}^{M} \log\left[\cosh(b_j + \sum_{i=1}^{N} W_{ij} s_i)\right]
\label{annpsi}
\end{equation} provies us the amplitude of a spin configuration, $S=(s_1, ...,s_N)$, where ${\bf x}=\{b_j, W_{ij}\}$ are the network parameters. The initial value of ${\bf x}$ is taken such that the manybody state is where each atom is in the ground state. To capture the dynamics, NQS are optimized such that the distance between the time-evolved wave functions $e^{-i\hat H\Delta t}\ket{\Psi(S, \bf x)}$ and $\ket{\Psi(S, \bf x + \dot{\bf x} dt)}$ given by the Fubini study metric,
\begin{equation}
        \mathcal{D}(\ket{\psi}, \ \ket{\phi})^2 = \arccos{\left(\sqrt{\frac{\langle\psi |\phi\rangle \langle\phi |\psi\rangle}{\langle\psi |\psi\rangle \langle\phi |\phi\rangle}}\right)}^2
    \end{equation}
 is minimized for each time step $\Delta t$. The above time-dependent variational principle transforms the time-dependent Schr\"odinger equation into a set of non-linear symplectic differential equations for the variational parameters \cite{sch20},
\begin{equation}\label{TDVP}
\dot{\bf x}(t) = -i \bf A^{-1}\bf F,
\label{xt}
\end{equation}
where ${\bf A}$ is a covariance matrix and $\bf F$ is the generalized force. The elements of the matrix $\bf A$ and the vector $\bf F$ are
\begin{eqnarray}
{\bf A}_{kk'} = \langle O^*_kO_{k'}\rangle - \langle O^*_k\rangle\langle O_{k'}\rangle\\
{\bf F}_k  = \langle O^*_kE_{\mathrm{loc}}\rangle - \langle O^*_k\rangle \langle E_{\mathrm{loc}}\rangle,
\end{eqnarray}
where $\langle ... \rangle$ is taken over $\Psi(S)$, and $O^*_{k'}(S)$ and $O_k(S)$ are variational derivatives given by
\begin{equation}
 O_k(S) = \frac{1}{\Psi(S)}\frac{\partial \Psi(S)}{\partial x_k}
\end{equation}
with $E_{\mathrm{loc}}$ being called the local energy, defined as
\begin{equation}
	E^S_{\mathrm{loc}}({\bf x}) = \frac{\langle S|\hat H|\Psi\rangle}{\langle S|\Psi\rangle}.
\end{equation}
 The equation~(\ref{TDVP}) is then solved using the adaptive Heun scheme.
%%%%%
\section{Excitation Dynamics}
\label{ssd}
%%%%%%%%
%%%%%%%%%%%%%%%%%%%%%%%%
%%Figure 1
%%%%%%%%%%%%%%%%%%%%%%%%
\begin{figure}
\centering
\includegraphics[width= 0.7\columnwidth]{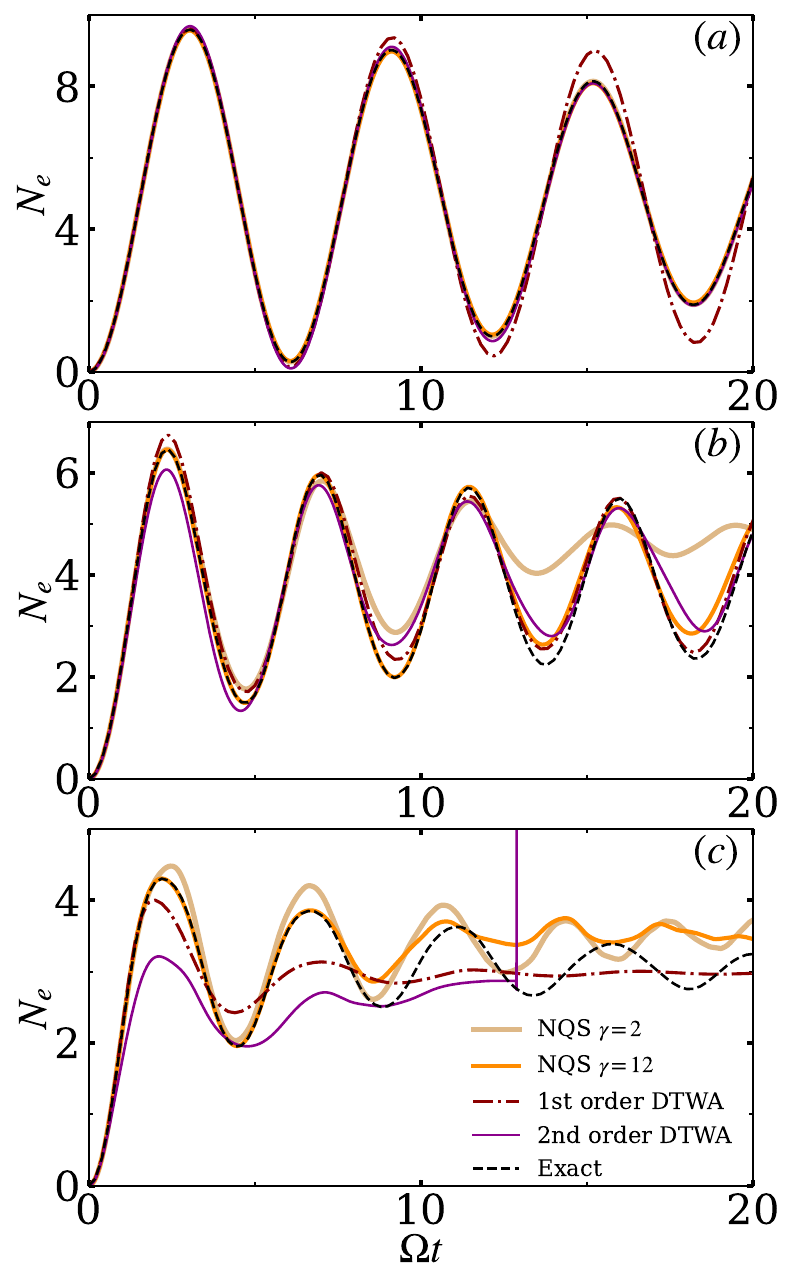}
\caption{The dynamics of the total number of excitations ($N_e$) in an atomic chain of $N=10$ for different $\tilde C_6$. We compare the exact results (dashed line) with first and second-order DTWA and NQS (solid lines). (a) is for $\tilde C_6=0.2\Omega$, (b) for $\tilde C_6=\Omega$ and (c) is for $\tilde C_6=3\Omega$. For $\tilde C_6=0.2\Omega$, both NQS and second-order DTWA are in excellent agreement with exact results, and as $\tilde C_6$ increases, they deviate at longer times. NQS is found to be more reliable at large $\tilde C_6$. Taking a larger $\gamma$ improved the NQS results for extended periods compared to smaller $\gamma$. In (c) for $\tilde C_6=3\Omega$, second-order DTWA experiences numerical instability at $\Omega t\approx 13$, and the larger the $\tilde C_6$ earlier the instability.}
\label{fig:1}
\end{figure}
%%%%%%%%%%%%%%%%
%%%%%%%%%%%%%%%%
%%%%%%%%%%%%%%%%%%%%%%%%
%%Figure 2
%%%%%%%%%%%%%%%%%%%%%%%%
\begin{figure}
\centering 
\includegraphics[width=1 \columnwidth]{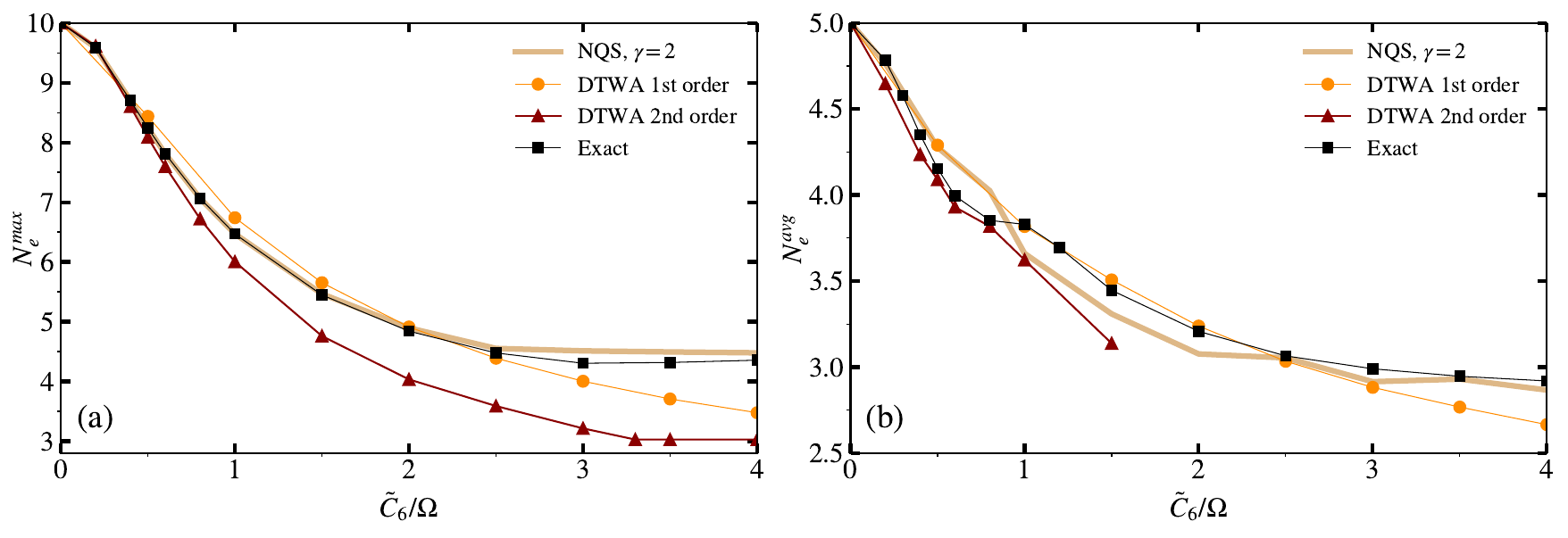}
\caption{(a) The maximum ($N_e^{max}$) and (b) average ($N_e^{avg}$) number of excitations as a function of $\tilde C_6$ in a 10 atom chain. We compare the exact results with that of DTWA and NQS. For $N_e^{avg}$, we take an average over a period of $\Omega T=100$ for exact results, whereas for DTWA first order and NQS, we took  $\Omega T=20$ and DTWA second order, we took  $\Omega T=10$. The numerical instability in the secon-order DTWA limits the averaging period to smaller values.}
\label{fig:2}
\end{figure}
%%%%%%
In the following, we compare the Rydberg excitation dynamics obtained from DTWA and NQS to the exact results for $N=10$. A ten-atom chain with rydberg admixed softcore interactions is experimentally studied in Ref. \cite{zei17}. The initial state is where all atoms are in their ground state. Figure~\ref{fig:1} shows the dynamics of the total number of excitations $N_e(t)$ for different values of $\tilde C_6=C_6/d^6$. Initially, $N_e=0$, and as time evolves, $N_e$ increases and eventually oscillates in time. The oscillation amplitude is larger in the first cycle and damps over the subsequent cycles. For small $\tilde C_6$ [see figure~\ref{fig:1}(a)], both NQS and second-order DTWA  show an excellent agreement with exact results for sufficiently long periods. $\gamma=2$ is sufficient to capture the dynamics as depicted in figure~\ref{fig:1}(a). The first-order DTWA deviates from the exact dynamics at longer times, indicating the importance of two-body correlations even at low $\tilde C_6$. 

As $\tilde C_6$ increases, the agreement with exact dynamics is lesser for both NQS and DTWA, especially at longer times, but qualitative features are captured. For instance, the damping of oscillation amplitude of  $N_e(t)$ is observed in both methods. The larger the $\tilde C_6$, the earlier the deviation from exact results occurs for both techniques. Better NQS results can be obtained by taking a larger $\gamma$, but increasing $\gamma$ will increase computational time. The effect of $\gamma$ on the dynamics is shown in \ref{gamma}.

As we have noticed, at larger $\tilde C_6$ [figure~\ref{fig:1}(c)], second-order DTWA shows more deviation than first-order. It also suffers from numerical instability and is found to depend on the interaction strength but not on the system size. We anticipate that the higher-order BBGKY terms are required to counter the instability for large interaction strengths ($\tilde C_6$). The latter are tedious to obtain, and numerical stability is generally not guaranteed. We also note that for larger $\tilde C_6$, oscillation amplitude shows a faster decay in DTWA results. We observe that NQS is more reliable if short-time dynamics is desired.       

%%%%%%%%%%%%%%%%%%
%%Ne average and maximum
%%%%%%%%%%%%%%%%%%

In figure~\ref{fig:2}, we show the dependence of maximum  ($N_e^{max}$), which is the amplitude of the first peak and average ($N_e^{avg}$) of $N_e$ on the interaction strength $\tilde C_6$. Both $N_e^{max}$ and $N_e^{avg}$ decrease with an increase in $\tilde C_6$ due to the blockade effect. For $\tilde C_6\ll\Omega$, we have $N_e^{max}\sim N$ and $N_e^{avg}\sim N/2$ and in a fully blockaded chain (limit $\tilde C_6\to \infty$) $N_e^{max}\sim 1$ and $N_e^{avg}\sim 1/2$.  For  $\tilde C_6>2\Omega$, in the exact results, $N_e^{max}\sim 5$ and $N_e^{avg}\sim 3$ saturate due to the nearest neighbor blockade, but the next nearest neighbor blockade requires an interaction strength of $\tilde C_6\sim 64 \Omega$. Therefore, as a function of $\tilde C_6$, different plateaus for $N_e^{max}$ and $N_e^{avg}$ would emerge, indicating the blockades of atoms at larger separations. Interestingly, as we showed in figure~\ref{fig:2},  $N_e^{max}$ and $N_e^{avg}$ are accurately captured by NQS with $\gamma=2$, even at larger $\tilde C_6$. Thus, regarding the average and peak value of the number of Rydberg excitations, a large $\gamma$ is not required, and NQS can be applied for even large interactions.

\begin{figure}
\centering
\includegraphics[width=0.9 \columnwidth]{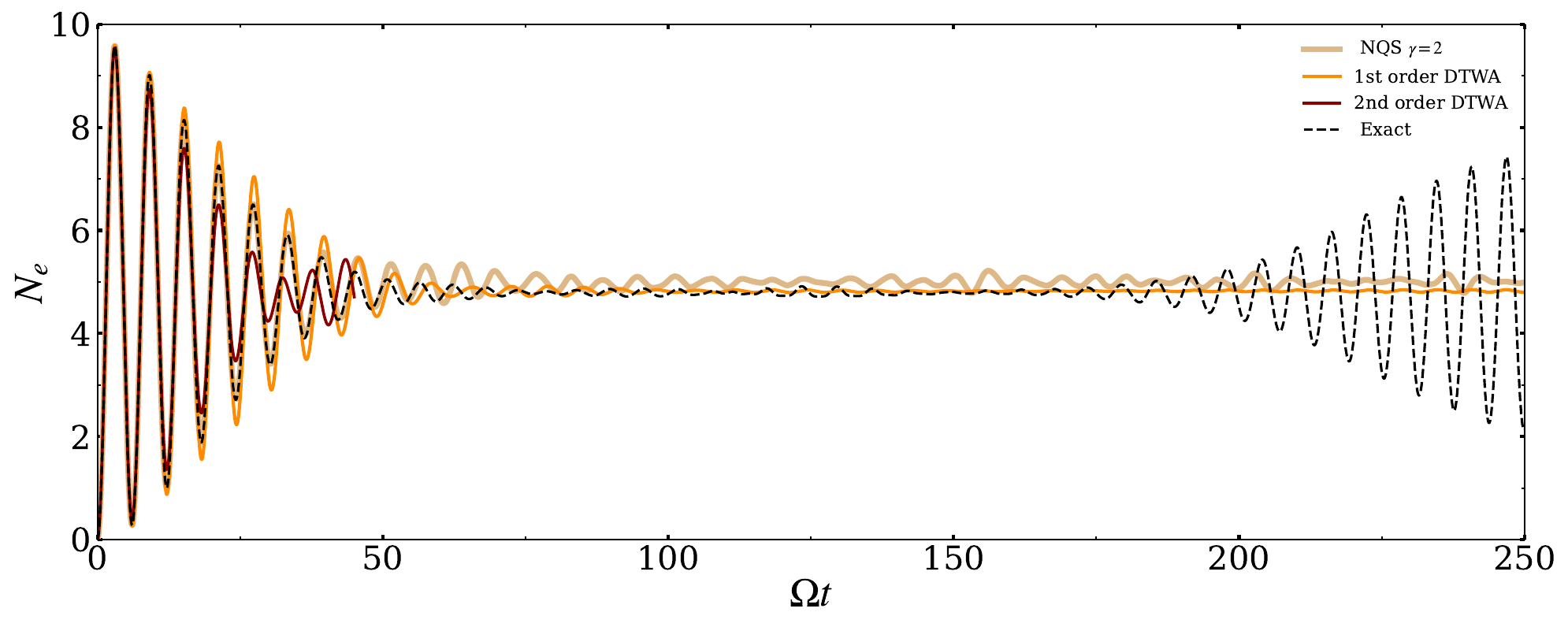}
\caption{The long-time dynamics of $N_e(t)$ for $N=10$ and $\tilde C_6=0.2\Omega$. We compare the exact results with that of first-order DTWA and NQS. Although both methods capture the collapse, the revival is absent. We also verified (but not shown) that increasing $\gamma$ does not affect the results for $\tilde C_6=0.2\Omega$, as it is small.}
\label{fig:3}
\end{figure}

%%%%%%%%%
For sufficiently small values of $\tilde C_6$ and at longer times, there exists collapse and revival of $N_e(t)$ about its average value \cite{wug15}, as shown in figure~\ref{fig:3}. A similar feature also arises in the population dynamics of a two-level atom coupled to a single-mode quantized light field in an optical cavity, described by the Jaynes-Cummings model (JCM). The collapse and revival in JCM are associated with the statistical and discrete nature of the photon field \cite{ebe80, rem87}. In the Rydberg chain, the collapse and revival in $N_e(t)$ can be attributed to the discrete nature (periodic lattice) of the system and interactions. DTWA and NQS capture the collapse of $N_e(t)$ to its average value, but the revival is not seen even after sufficiently long times [see figure~\ref{fig:3}]. We also verified (but not shown) that increasing $\gamma$ does not change the results shown in figure~\ref{fig:3} for $\tilde C_6=0.2$.

%%%%%%%%%%%%%%%%%%%%%%%%
%%Figure 3
%%%%%%%%%%%%%%%%%%%%%%%%

%%%%%%%%%%%%%%%%%%%%%%%%%%%%%%%
\section{Correlation dynamics}
\label{re}
%
%In the following we analyze the dynamics of the various correlation functions. 
%\subsection{Two-point correlation function}
%First, we look at the two-point correlations of the excitation dynamics
%\begin{eqnarray}
   % C(d,t) &= \sum_i \left( \braket{\hat{\sigma}_{ee}^i\hat{\sigma}_{ee}^{i+d}} - \braket{\hat{\sigma}_{ee}^i}\braket{\hat{\sigma}_{ee}^{i+d}}\right) \nonumber \\
    %&= \sum_i \frac{1}{4} \left( \braket{\hat{\sigma}^{i}_z\hat{\sigma}^{i+d}_z} - \braket{\hat{\sigma}^{i}_z}\braket{\hat{\sigma}^{i+d}_z} \right).
%\end{eqnarray}

\subsection{Bipartite R\'enyi entropy}

%%%%%%%%%%%%%%%%%%%%%%%%
%%Figure 4
%%%%%%%%%%%%%%%%%%%%%%%%
\begin{figure}
\centering
\includegraphics[width=0.8 \columnwidth]{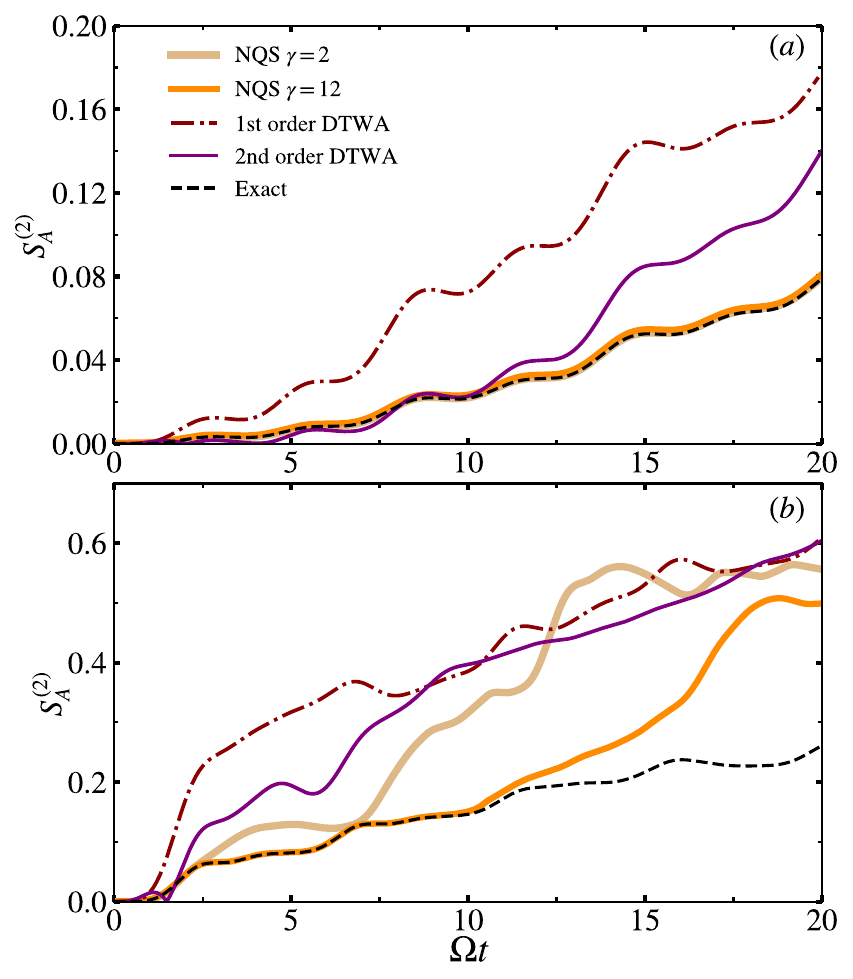}
\caption{The dynamics of bipartite R\'enyi entanglement entropy for different (a) $\tilde C_6=0.2$ and (b) $\tilde C_6=1$. We compare the exact results with that of DTWA and NQS. For $\tilde C_6=0.2$, $\gamma=2$ is sufficient enough to capture the results, whereas for $\tilde C_6=\Omega$, $\gamma=12$ gives a better agreement with the exact results.  }
\label{fig:4}
\end{figure}

%%%%%%%%%%%%%%%%%%%%%%%%
%%Figure 5
%%%%%%%%%%%%%%%%%%%%%%%%
\begin{figure}
\centering
\includegraphics[width=0.8 \columnwidth]{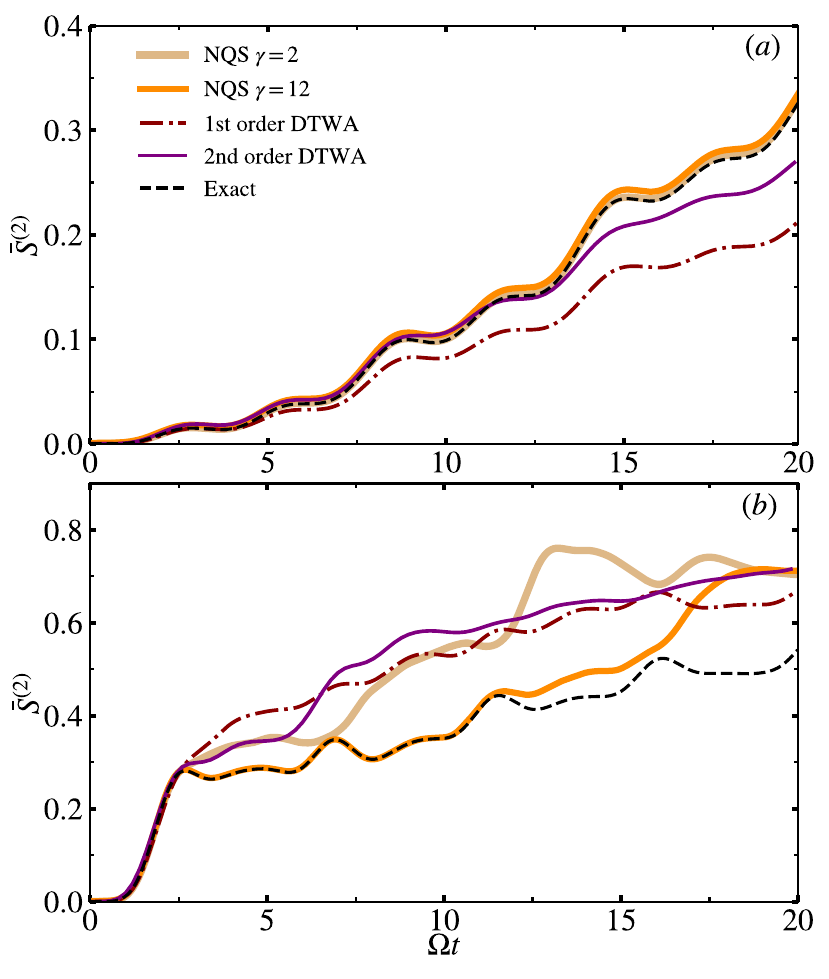}
\caption{The dynamics of the mean two-site entropy for (a) $\tilde C_6=0.2$ and (b) $\tilde C_6=1$. We compare the exact results with that of DTWA and NQS.  For $\tilde C_6=0.2$, $\gamma=2$ is sufficient enough to capture the results, whereas for $\tilde C_6=\Omega$, $\gamma=12$ gives a better agreement with the exact results. }
\label{fig:5}
\end{figure}

The quantum correlations in a many-body system are of significant importance, and here, we characterize them using the second-order R\'enyi entropy. The second-order R\'enyi entropy is defined as,
\begin{equation}
 S^{(2)}_A (t)= -\frac{1}{N_A}\log({\rm Tr}[(\hat{\rho}_A(t)^2)]),
 \label{sae}
\end{equation}
where $N_A$ is the number of atoms in the subsystem $A$ and the reduced density matrix for the subsystem A is
\begin{equation} \label{rdm}
\hat{\rho}_A(t)=\sum_{\alpha} \frac{1}{N_s}\hat{\mathcal A}^{\vec\alpha_A}_{1...N_A}(t),
\end{equation}
where $\vec \alpha_A=\{\alpha_1, \alpha_2, ...,\alpha_{N_A}\}$. In first-order DTWA, $\hat{\mathcal A}^{\alpha_1...\alpha_{N_A}}_{1...N_A}=\prod_{i \in A} \hat{\mathcal{A}}_i$ and the R\'enyi entropy takes the simple form \cite{kun21}, 
\begin{equation}
S^{(2)}_A (t) = -\frac{1}{N_A}\log\left[\sum_{\alpha,\alpha'}\frac{1}{N_s^2} \prod_{i \in A} \frac{1}{2} \left( \mathcal{I} + \sum_{\mu} a_i^{\mu}(t)a_i^{\prime\mu}(t) \right) \right],
\label{s2}
\end{equation}
where $a_i(0) = {\bf r}_{\alpha_i}$ and  $a'_i(0) = {\bf r}_{\alpha'_i}$ are the initial conditions, which are evolved independently. In the second-order DTWA, additional correlation-dependent terms appear in equation~(\ref{s2}). The computation of $S^{(2)}_A (t)$ in equation~(\ref{s2}) involves a phase-space average weighted with two initial Wigner functions. In the case of NQS, the reduced density matrix can be directly obtained using the QuTiP, and then, we compute $S^{(2)}_A (t)$ using equation~(\ref{sae}). We also calculate the mean two-site entropy, 
\begin{equation}
    \bar{S}^{(2)}(t) = \frac{1}{N - 1}\sum_{\langle i j\rangle} S^{(2)}_{ij}(t)
\end{equation}
Here, $S^{(2)}_{ij}$ is the 2nd order R\'enyi entanglement entropy for the subsystem containing two neighboring sites. Previously, it has been employed to study many-body localization and thermalization in a one-dimensional Heisenberg model \cite{ace17}.

The dynamics of bipartite $S^{(2)}_A (t)$ for $N=10$ or $N_A=5$ is shown in figure~\ref{fig:4}. At $t=0$, we have $S^{(2)}_A (t)=0$ since the initial state is a product state where each atom is in $|g\rangle$. As time progresses, correlations build in the system, and $S^{(2)}_A (t)$ grows. Larger the $\tilde C_6$, faster the growth of the correlations.  Even though first-order DTWA captured the population dynamics accurately for small $\tilde C_6$ [see figure~\ref{fig:1}(a)], it overestimates the bipartite R\'enyi entropy [see figure~\ref{fig:4}(a)]. The results from second-order DTWA shows a better agreement compared to that of first order DTWA. In contrast,  NQS captures the correlation dynamics accurately for small interactions even with $\gamma=2$. When the RRI is increased, to $\tilde C_6=\Omega$, $\gamma=2$ is not sufficient to determine the dynamics of the correlations accurately. Using $\gamma=12$, the dynamics accurately obtained upto a duration of $\Omega t\approx 10$. 

\begin{figure}
\centering
\includegraphics[width=1 \columnwidth]{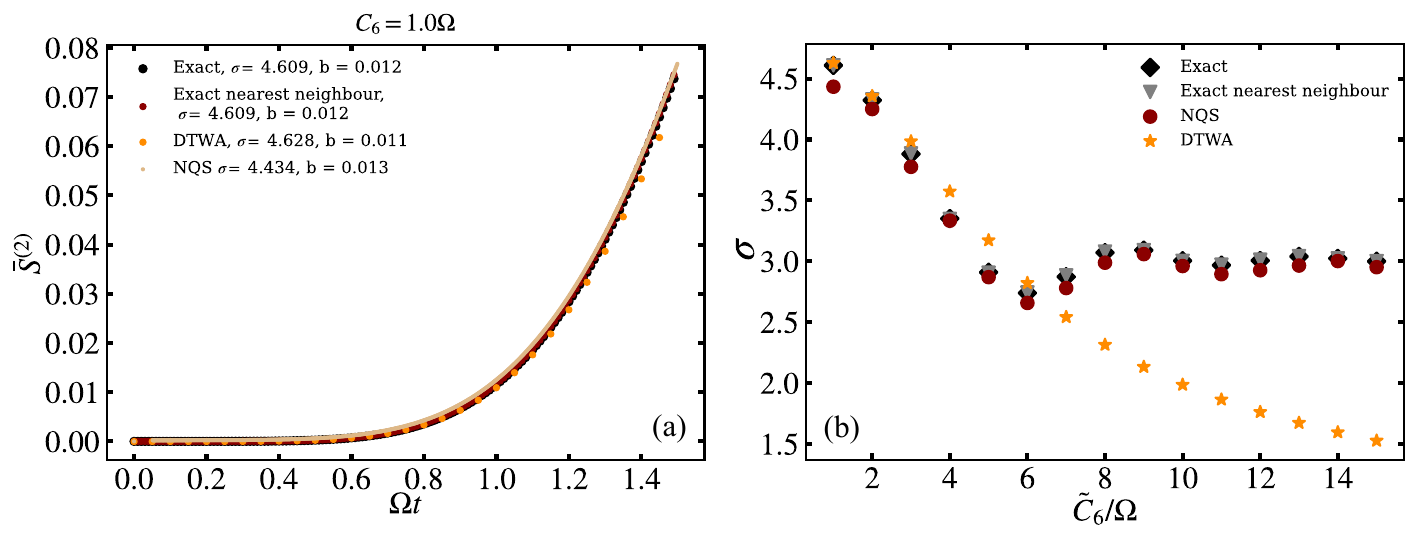}
\caption{(a) The initial power-law growth of mean two-site R\'enyi entropy in a 10 atom chain for $\tilde C_6=\Omega$. The results from DTWA and NQS are in good agreement with the exact results. (b) The exponent $\sigma$ as a function of $\tilde C_6/\Omega$ for $N=10$. The NQS results are in an excellent agreement with that of exact results. Since we look at the initial correlation growth, the long-range interactions do not play any role and nearest neighbor model gives identical behavior.}
\label{fig:6}
\end{figure}

Interestingly, the dynamics of mean two-site entropy [see figure~\ref{fig:5}] obtained from all methods show better agreement with each other even for $\tilde C_6=\Omega$. We anticipate that taking an additional average over all nearest neighbor sites smoothens out the fluctuations and makes the DTWA results closer to the exact ones. In general, we observed two stages in the correlation growth: faster power-law growth in the initial stage and slower growth at longer times, as shown in figure~\ref{fig:5}(b). For the initial growth, we see that the mean two-site entropy behaves as $\bar{S}^{(2)}(t)\approx b (\Omega t)^\sigma$ [see figure~\ref{fig:6}(a)], where $b$ and $\sigma$ are interaction dependent parameters. For $\tilde C_6=\Omega$, we obtain $b=0.012$ and $\sigma=4.609$, and DTWA and NQS accurately capture them. We verified that the coefficient $b$ and exponent $\sigma$ do not show any dependence on the system size up to $N=18$ (exact results) and decrease with an increase in $\tilde C_6$ before they saturates to a value of $b\sim 0.1$ (not shown) and $\sigma\sim 3$ [see figure~\ref{fig:6}(b)]. We anticipate that when the nearest neighbor Rydberg blockade pitches in ($\tilde C_6\approx 64\Omega$), $\sigma$ will again decrease and eventually saturate, leading to a staircase-like structure. Since long-range interactions do not play a significant role in the initial correlation growth, truncating the RRIs to the nearest neighbor also leads to the same results as shown in figure~\ref{fig:6}.
 
 %%%%%%%%%%%%%%%%%%%%%%%%
%%Figure 2
%%%%%%%%%%%%%%%%%%%%%%%%
%%%%%%
%%%%%%%%%%%%%%%%%%%%%%%%%%%%%%%
\section{Conclusions and Outlook}
\label{con}
As the demand for numerical methods to capture the dynamics of quantum many-body systems is high, we benchmark DTWA and NQS based on RBM-like architecture
using the exact results for the excitation and second-order R\'enyi entropy dynamics in a chain of ten two-level Rydberg atoms. In DTWA, we go beyond the mean-field approximation and include the two-body correlations via BBGKY hierarchy, which is tedious. DTWA and NQS results agree with the exact results for weak Rydberg-Rydberg interactions for relatively more extended periods. NQS results generally improve by taking many hidden spins, but it increases computational time. The failure of the DTWA second-order method at longer periods and strong interactions indicates that multi-particle correlations are crucial for predicting the dynamics accurately. Although the population dynamics at longer times showed deviations for large interaction strengths, the average number of excitations calculated using NQS shows excellent agreement with the exact results, even using a relatively small number of hidden spins or neurons. While analyzing the second-order R\'enyi entropy, we established a power law behavior in its initial growth, with an exponent depending significantly on the interactions and independent of the system size. Considering the short time scales involved in the experiments and the initial product states, we conclude that NQS would help analyze the dynamics in a chain of interacting Rydberg atoms. 

We would like to see how these methods can be modified so that large interactions can be considered. Further, we would like to understand the instabilities in DTWA-BBGKY formalism. The same analysis can be extended to multi-dimensional setups, where the first-order DTWA, being a mean field, would work better and also to consider time-dependent Hamiltonians \cite{blu21, bas18, nir20, mal21, dhi23}. In particular, to analyze the formation of many-body states under quantum quenching, for instance, the dynamical crystallization \cite{poh10}. Also, the spontaneous emission from the Rydberg state will be incorporated to study the dissipative dynamics.  

%%%%%
\section{Acknowledgments}
\label{ack}
We acknowledge UKIERI-UGC Thematic Partnership No. IND/CONT/G/16-17/73 UKIERI-UGC project. W.L further thanks the support from the EPSRC through Grant No. EP/R04340X/1 via the QuantERA project "ERyQSenS" and the Royal Society through the International Exchanges Cost Share award No. IEC$\backslash$NSFC$\backslash$181078.. R.N. further acknowledges DST-SERB for Swarnajayanti fellowship File No. SB/SJF/2020-21/19, and the MATRICS grant (MTR/2022/000454) from SERB, Government of India, and National Supercomputing Mission (NSM) for providing computing resources of 'PARAM Brahma' at IISER Pune, which is implemented by C-DAC and supported by the Ministry of Electronics and Information Technology (MeitY) and Department of Science and Technology (DST), Government of India. V.N. acknowledges the funding from DST India through an INSPIRE scholarship. We also acknowledge funding from National Mission on Interdisciplinary Cyber-Physical Systems (NM-ICPS) of the Department of Science and Technology, Govt. Of India through the I-HUB Quantum Technology Foundation, Pune INDIA. In particular, V.S acknowledges the Chanakya fellowship awarded by I-HUB Quantum Technology Foundation, Pune INDIA. Finally, we acknowledge QuSpin \cite{wei17, wei19}, QuTiP  \cite{joh12, joh13}, and jVMC packages \cite{sch22a, sch22b}. 

%%%%%%%%%
\appendix
\section{Effect of $\gamma$ on the Excitation dynamics}

\label{gamma}
\begin{figure}
\centering
\includegraphics[width= 0.7\columnwidth]{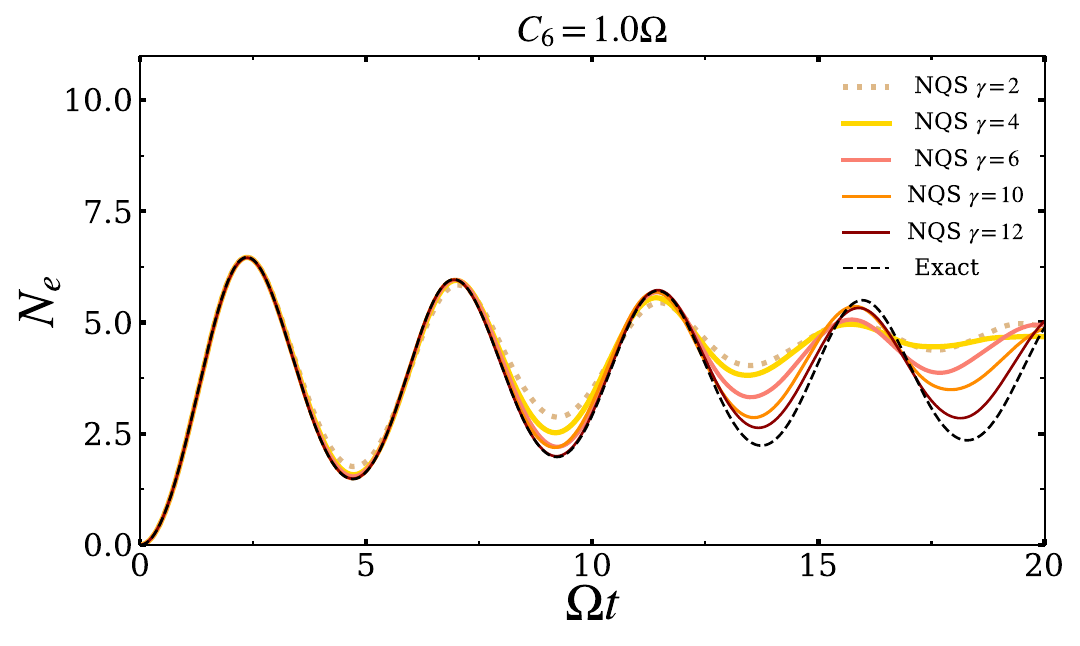}
\caption{The NQS dynamics of the total number of excitations ($N_e$) in an atomic chain of $N=10$ for  $\tilde C_6=\Omega$ and different number of hidden neurons ($\gamma$). }
\label{fig:a1}
\end{figure}

In figure~\ref{fig:a1}, we show the excitation dynamics obtained via NQS in a ten-atom chain for different numbers of hidden neurons, i.e., by varying $\gamma$ and $\tilde C_6=\Omega$. Though the results have been improved with an increase in $\gamma$, the deviation at longer times persists. Considering the experimental studies involve shorter periods, NQS will be helpful, and relatively long-time dynamics can be captured with weaker interactions.
%%%%%%%%%%%%%%%%%%%%%%%%
%%Figure 1
%%%%%%%%%%%%%%%%%%%%%%%%
%%%%%%%%%%%%%%%%

\providecommand{\newblock}{}

\end{document}